\documentclass[12pt,preprint]{aastex}
\def\({\left(}
\def\){\right)}
\def\[{\left[}
\def\]{\right]}

\usepackage{graphicx}
\usepackage{natbib}
\usepackage[usenames,dvips]{color}

\begin{document}
\title{ The effect of coronal radiation on a residual inner disk in the
low/hard spectral state of black hole X-ray binary systems}

\author{B. F. Liu\altaffilmark{1,*}
\email{bfliu@ynao.ac.cn}
C. Done\altaffilmark{2}
\and
Ronald E. Taam\altaffilmark{3,4}
\email{r-taam@northwestern.edu}}
\altaffiltext{1}{National Astronomical Observatories/Yunnan Observatory,
Chinese Academy of Sciences, P.O. Box 110, Kunming 650011, China}
\altaffiltext{2}{Department of Physics, University of Durham, South Road, Durham, DH1 3LE, UK}
\altaffiltext{3}{Northwestern University, Department of Physics and Astronomy,
2131 Tech Drive, Evanston, IL 60208}
\altaffiltext{4}{Academia Sinica Institute of Astronomy and Astrophysics - TIARA,
P.O. Box 23-141, Taipei, 10617 Taiwan}
\altaffiltext{*}{Present Address: National Astronomical Observatories, Chinese Academy of Sciences, Beijing 100012, China. bfliu@nao.cas.cn}

\begin{abstract}

Thermal conduction between a cool accretion disk and a hot inner
corona can result in either evaporation of the disk or condensation of
the hot corona. At low mass accretion rates, evaporation dominates and
can completely remove the inner disk. At higher mass accretion rates,
condensation becomes more efficient in the very inner regions, so that
part of the mass accretes via a weak (initially formed) inner disk
 which is separated from the outer disk by a fully
evaporated region at mid radii.  At still higher mass accretion rates,
condensation dominates everywhere, so there is a continuous cool disk
extending to the  innermost stable circular orbit.  We extend these
calculations by including the effect of irradiation by the hot corona on the
disk structure. The flux which is not reflected is reprocessed in the disk,
adding to the intrinsic thermal emission from gravitational energy release. This
increases the seed photons for Compton cooling of the hot corona, enhancing
condensation of the hot flow and re-inforcing the residual inner disk
rather than evaporating it. Our calculations confirm that a
residual inner disk can co-exist with a hard, coronally dominated,
spectrum over a range of $0.006<\dot m<0.016$ (for $\alpha=0.2$).
This provides an explanation for the weak thermal component seen
recently in the low/hard state of black hole X-ray binary systems.
\end{abstract}

\section{Introduction}

It is well known that black hole X-ray binaries (BHXRBs) exhibit
various spectral states.  As recently reviewed by Remillard \&
McClintock (2006), such phenomena are quite common in BHXRBs. It is
generally accepted that the transition from a spectrally soft to hard
state is caused by a change in the accretion mode in the inner disk
regions from a cool, geometrically thin structure at high accretion
rates to a hot, geometrically thick, radiatively inefficient accretion
flow (RIAF/ADAF) at low accretion rates.

Many theoretical studies have attempted to identify the mechanism
responsible for this transition, taking into account processes such as radial
conductive energy transport (Honma 1996; Manmoto \& Kato 2000) and vertical
 evaporation (Meyer et al. 2000a; Rozanska \& Czerny 2000; Spruit \& Deufel 2002;
Dullemond \& Spruit 2005; Mayer \& Pringle 2007; Bradley \& Frank
2009).  Of these, the disk evaporation model is one of the more
promising. Thermal conduction between a cool disk and a hot corona can
result in either evaporation of the cold disk or condensation of the
hot corona (Meyer et al. 2000a,b; Liu et al. 2002, Liu et
al. 2006, Meyer et al. 2007). At accretion
rates  higher than the maximal evaporation rate, $\dot m\ga 0.03$ (scaled by the Eddington rate $\dot M_{\rm
Edd}=1.39\times 10^{18}M/M_\odot$ g s$^{-1}$), for
the specific case of $\alpha=0.2$, a full disk
coexists with a corona. At very low accretion rates, $\dot m\la
0.006$, the inner disk is completely evaporated into an ADAF. This
gives an interpretive framework for features observed in BHRXBs,
such as disk truncation  as the cause of the spectral state transition (Liu
et al. 1999; Meyer et al. 2000b; Liu \& Meyer-Hofmeister 2001) and luminosity
hysteresis between the hard-to-soft and soft-to-hard transitions
(Meyer-Hofmeister et al. 2005; Liu et al.  2005; Bradley \& Frank
2009).

It also predicts more complex behaviour at intermediate accretion
rates, $0.006\la\dot m\la 0.03$. Here the disk is truncated into a
composite form with a coronal gap settling between an inner disk and
an outer disk, with the inner disk carrying a fraction of the mass
accretion rate, covered by a corona which carries the remaining
material (see, e.g., fig.1 in Meyer-Hofmeister, Liu, Meyer 2009).  This
could explain the complex spectra seen in the intermediate state (Liu
et al. 2006; Meyer et al. 2007), immediately following the soft
state and predicts that a weak inner cool disk component can co-exist
with a coronally dominated spectrum (Liu et al. 2007; Taam et al. 2008).

This residual inner disk provides a potential explanation of the
recent observation of a weak thermal component in some BHXRBs in
the low/hard state (Miller et al. 2006a,b; Tomsick et al. 2008, Reis
et al 2009; 2010; Chiang et al 2010). However, the calculations to
date have only studied evaporation as a means of coupling the disk and
corona, whereas radiative coupling from irradiation of the disk by the
corona is also important.
The corona illuminates the disk, and most of this irradiation
is reprocessed, enhancing the luminosity of the disk, and these
photons provide additional Compton cooling of the corona so there is
strong feedback between the disk and corona (Haardt \& Maraschi
1993). Here, we investigate the effect of this additional radiative
coupling between the disk and corona, especially to determine
whether the inner disk can survive or is completely evaporated by
irradiation.

In \S 2 we briefly introduce the disk corona evaporation/condensation
model and its extension to incorporate the effect of coronal
irradiation. Numerical results from these models are presented in \S
3, showing the strength of inner disk relative to the corona/ADAF.
Our conclusion is presented in \S 4.

\section{The model for an inner disk around a black hole including irradiation}

The disk/corona model adopted here is based on the studies of Liu et
al. (2006) and Meyer et al. (2007) with modifications recently
incorporated in the works of Liu et al. (2007) and Taam et
al. (2008). It is assumed that an ADAF-like corona (given by the
self-similar equations of Narayan \& Yi 1995, with $\alpha=0.2$ and $\beta=0.8$)
lies above a thin disk. The corona
is heated by the viscous release of gravitational energy of accreted
gas and dominantly cooled by vertical conduction and Compton
scattering of soft photons emitted by the underlying disk.
In the vertical transition layer between
the disk and corona, an equilibrium is established between the
conductive flux from the upper corona, bremsstrahlung radiation, and
vertical enthalpy flux. For a given distance from the black hole, a
fraction of the disk gas is heated and evaporated to the corona when
the conduction flux is too large to be radiated away. On the other
hand, a certain amount of coronal gas is cooled down, condensing to
the disk if the bremsstrahlung radiation is more efficient than the
conduction.  At accretion rates around a few percent of Eddington value, gas evaporates from the disk to the corona around a few hundred Schwarzschild radii and partially condenses back to the disk in the innermost region (Liu et al. 2006; 2007; Meyer et al. 2007).

\subsection{The condensation feature without irradiation}
 In the case of conduction dominant cooling the condensation rate is given as (Liu et al. 2006; 2007; Meyer et al. 2007; Taam et al. 2008)
\begin{equation}\label{condensation-cnd}
\dot m_{\rm cnd}(r)= 3.23\times 10^{-3}\alpha^{-7}\dot m^3 \[1-6(r/r_d)^{1/2}+5(r/r_d)^{3/5}\],
\end{equation}
where $r_d$ is the critical radius from which gas flowing changes from evaporation in the outer region to condensation in the inner region. Here, $r_d$, is expressed as
\begin{equation}\label{r-cnd1}
r_{d}=0.815\alpha^{-28/3}\dot m^{8/3}.
\end{equation}
In the case of Compton dominant cooling the condensation rate is
\begin{equation}\label{condensation-cmp}
\dot m_{\rm cnd}(r)=A\left\{2B\[\({r_{\rm d}\over r}\)^{1/2}-1\]-\int_{r/3}^{r_{\rm d}/3}x^{1/5}\(1-x^{-1/2}\)^{-2/5}dx \right\},
\end{equation}
where
\begin{equation}\label{A}
A=6.164 \times 10^{-3}\alpha^{-7/5}m^{-2/5}\dot m^{7/5}\({T_{\rm eff, max}\over 0.3keV}\)^{-8/5}
\end{equation}
\begin{equation}\label{B}
B=3.001\alpha^{ -14/15}m^{2/5}\dot m^{ 4/15}\({T_{\rm eff, max}\over 0.3keV}\)^{8/5}\({r\over 3}\)^{1/2},
\end{equation}
and $r_d$ is the critical condensation radius,
\begin{equation}\label{r-cnd2}
r_{\rm d}\[1-\({3\over r_d}\)^{1/2}\]^{-4/7}=14.417\alpha^{-4/3}m^{4/7}\dot m^{8/21}\({T_{\rm eff,max}\over 0.3keV}\)^{16/7},
\end{equation}
$T_{\rm eff, max}$ is the maximal disk temperature, which depends on the condensation rate integrated from outer edge of the inner disk ($r_d$) to the distance $r_{\rm tmax}=49/12$
\begin{equation}\label{Teff-max}
T_{\rm eff, max}=1.3348\times 10^7  m^{-1/4}\dot m_{\rm cnd}^{1/4}(r_{\rm tmax}){\rm K}.
\end{equation}
The radial distribution of disk temperature is approximately expressed as,
\begin{equation}
T_{\rm eff}(r)=2.05T_{\rm eff,max}\({3\over r}\)^{3/4}\[1-\({3\over r}\)^{1/2}\]^{1/4}.
\end{equation}

For a black hole with given mass $m$ (in unit of solar mass), accretion rate $\dot m$ and viscosity parameter $\alpha$, the condensation rate can be determined by Eq.\ref{condensation-cnd} or Eq.\ref{condensation-cmp}.
In the Compton dominant case, an effective  temperature ($T_{\rm eff, max}$) is presumed for calculating the condensation rate, with which a new temperature is derived from Eq.\ref{Teff-max}. Iterative calculations
are carried out until the presumed temperature is consistent with the derived value.
Compton cooling becomes dominant when the accretion rate ($\dot m$) is high or/and the disk is heated to a high temperature (for details see Liu et al. 2007). At low accretion rates, Compton cooling is either negligible or only dominant in a very narrow inner region. In this case, the total condensation rate (and hence the disk temperature) is an integral over the conduction dominant region and Compton dominant region (if exist), and the size of the inner disk is determined by Eq.\ref{r-cnd1}.

\subsection{Inclusion of irradiation}
Since the inner disk is optically thick, the irradiating photons are
primarily reprocessed as blackbody radiation.  These additional soft
photons propagate through the corona and lead to more efficient
Compton scatterings. This leads to enhanced coronal cooling, so a
greater amount of coronal gas condenses into the inner
disk. Therefore, the effect of the irradiation is to increase the
condensation, building up a relatively strong inner disk. This
increase is dependent on the irradiative luminosity, the
absorption/scattering of the associated photons by the inner disk, and
the covering factor of the inner disk as seen from the ADAF.  The
value of the albedo is usually taken as a constant equal to 0.15
(Haardt \& Maraschi 1993). Both the irradiation luminosity and
covering factor are determined by the accretion rate, the latter
dependence resulting from the dependence of the inner disk size on the
accretion rate.

\subsubsection{The covering factor of a disk to an ADAF}\label{covering}

Assuming that the radiation from the ADAF radiation can be represented
as a point source lying above the disk center at a height of $H_s$,
the covering factor of a disk ring at distances between $R_{\rm in}\le
R\le R_{\rm out}$ is given by
\begin{equation}\label{coverf}
f=\int_{R_{\rm in}}^{R_{\rm out}}{H_s\over 4\pi (R^2+H_s^2)^{3/2}} 2\pi R dR={1\over 2}\[{1+\({R_{\rm in}
\over H_s}\)^2}\]^{-{1\over 2}}-{1\over 2}\[{1+\({R_{\rm out}\over H_s}\)^2}\]^{-{1\over 2}}.
\end{equation}
For a disk characterized by $R_{\rm in}=0$ and $R_{\rm out}=\infty$,
Eq.\ref{coverf} yields $f=1/2$, indicating that half of the ADAF
radiation is intercepted by a full disk.  For a small disk ring
characterized by $3R_s\le R\le 10R_s$, the covering factor is 0.2,
provided that the height of the irradiation source is, $H_s=R_{\rm
in}=3R_s$ ($R_S$ is the Schwarzschild radius).  Therefore, only a
small fraction of the ADAF radiation illuminates the truncated inner
disk ring.

\subsubsection{The irradiations from the corona}\label{irradiation}
Assuming that the total intrinsic luminosity of the corona above and
below the disk is given as $L_{\rm c,in}$, the illumination flux to the disk
surface per unit area is,
\begin{equation}\label{F_ir0}
F_{\rm ir}={1\over 2}L_{\rm c,in}(1-a){H_s\over 4\pi (R^2+H_s^2)^{3/2}},
\end{equation}
where a fraction of the irradiation flux, $a$,  is assumed to be reflected at the disk surface.  To
simplify the calculation of the condensation/evaporation rate, Eq.\ref{F_ir0} is approximated as,
\begin{equation}\label{F_ir}
F_{\rm ir}\approx {3L_{\rm c,in} (1-a) \over 8\pi R^3 }H_s\[1-\({3R_s\over R}\)^{1/2}\],
\end{equation}
which is equivalent to assuming the covering factor per unit surface area, ${H_s\over 4\pi
(R^2+H_s^2)^{3/2}}={3 H_s\over 4\pi R^3 }\[1-\({3R_s\over R}\)^{1/2}\]$. The coefficient of 3 in
Eq.(\ref{F_ir}) is a normalization factor determined by the fact that a full disk covers half of the
sky of a corona as shown in \S\ref{covering} (that is, $f\approx \int_{3R_s}^{\infty}{3 H_s\over 4\pi R^3}
\[1-\({3R_s\over R}\)^{1/2}\]2\pi R dR=1/2$).   The expression for $F_{ir}$ given by eq.(\ref{F_ir}) is close to the exact expression for
the irradiating flux given by eq.(\ref{F_ir0}) at $R=(49/36)R_s$ where a standard disk reaches its maximal
temperature and, hence, is a good approximation.

The illumination flux heats the optically thick disk and is eventually
re-emitted as soft photons.  Thus, the radiative flux from the disk is
composed of both accretion released energy and irradiative energy by
the corona, which we express as
\begin{equation}\label{T_eff}
F_{\rm d}=\sigma T_{\rm eff}^4={3GM\dot M_{\rm cnd} \[1-\({3R_s\over R}\)^{1/2}\]\over 8\pi R^3}
\[{1+{6L_{\rm c,in}(1-a)\over \dot M_{\rm cnd}c^2}}{H_s\over 3R_s}\].
\end{equation}
 The corresponding disk temperature is modified by irradiation as,
\begin{equation}
T_{\rm eff}(r)=2.05T_{\rm eff,max}\({3\over r}\)^{3/4}\[1-\({3\over r}\)^{1/2}\]^{1/4}
\[{1+{6L_{\rm c,in}(1-a)\over \dot M_{\rm cnd}c^2}}{H_s\over 3R_s}\]^{1/4}.
\end{equation}
Therefore, the disk temperature is raised by a factor of $\[1+{6L_{\rm c,in}(1-a)\over \dot M_{\rm cnd}c^2}{H_s\over 3R_s}\]^{1/4}$. This factor is added to the expression of $T_{\rm eff,max}$ (Eq.\ref{Teff-max}) in calculating the condensation rate with Eqs.\ref{condensation-cmp}-\ref{r-cnd2}.

\section{Results}
Given the mass of a black hole and an accretion rate, the condensation
rate, corona luminosity and size of an inner disk can be calculated.
From these quantities, the covering factor and, hence, the
irradiating flux at any given distance can also be determined. As the
density of the soft photons,
representing reprocessed irradiating photons and photons originating from the viscous dissipation in the disk,
differs from the unirradiated case, iterative calculations are
performed until the derived soft photon density is consistent with the
presumed one.
For $M=10M_\odot$, $\alpha=0.2$,
$H_s=3R_s$, $a=0.15$, $\dot
m\equiv\dot M/\dot M_{\rm Edd}< 0.03$
 (so that a composite corona/ADAF geometry can form) we calculate the
condensation of this coronal flow to an inner, residual disk and
determine the relative strength of the corona with respect to the
disk.

\begin{figure}
\plotone{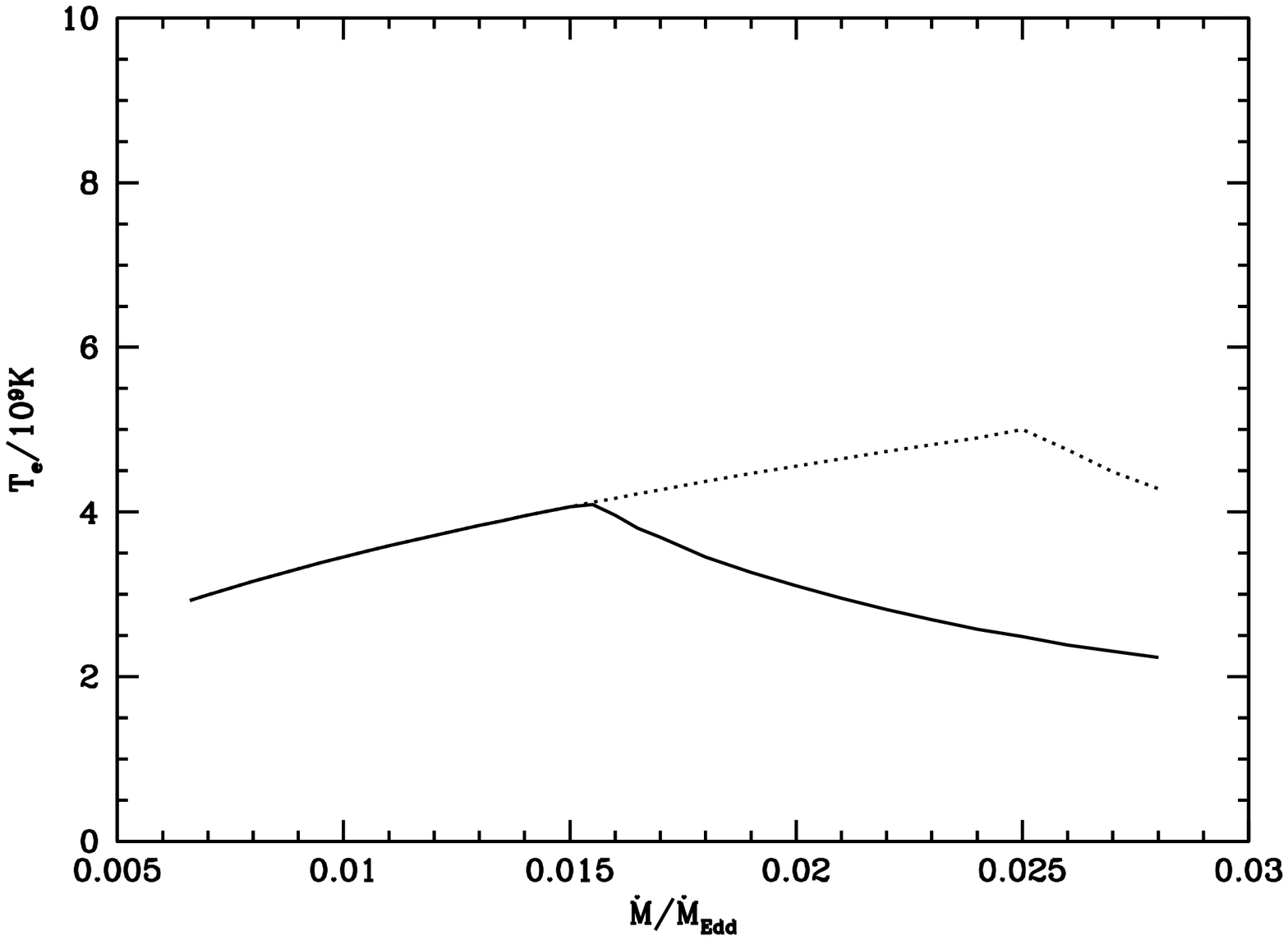}
\caption{\label{f:Te} The variation of the electron temperature in the corona with respect to the mass
accretion rate for the cases with irradiation (solid line) and
without irradiation (dotted line). Irradiation does not affect the temperature at low accretion rates, while it leads
to lower temperatures at higher accretion rates. The maximum temperature corresponds to the transition
where the dominant cooling process changes from conduction at low accretion rates to Compton scattering
at high accretion rates.}
\end{figure}

\begin{figure}
\plotone{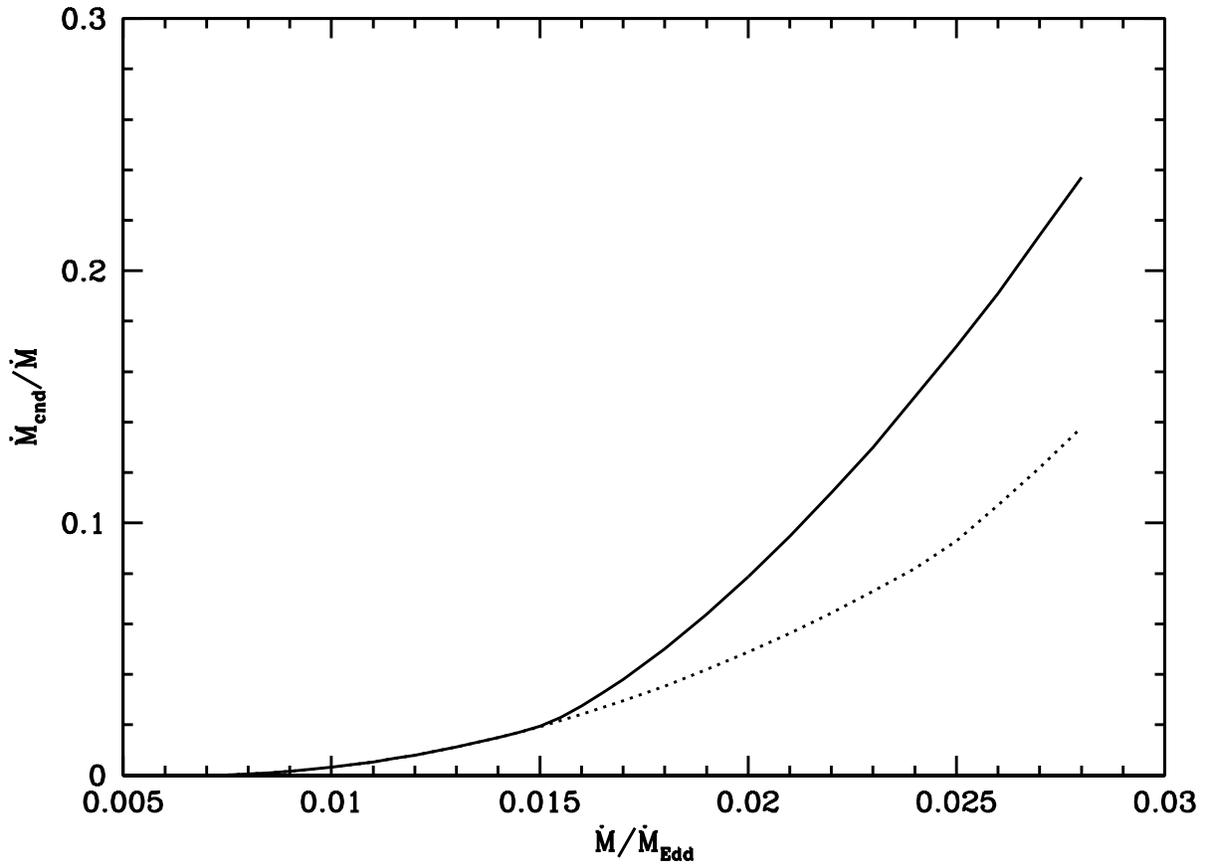}
\caption{\label{f:condensation} Comparison of the condensation fraction of the coronal accretion flow for
the cases with irradiation (solid line) and without irradiation (dotted line). The condensation fraction
is unaffected by the irradiation at low accretion rates, while it is increased by the irradiation at high accretion rates.}
\end{figure}

\begin{figure}
\plotone{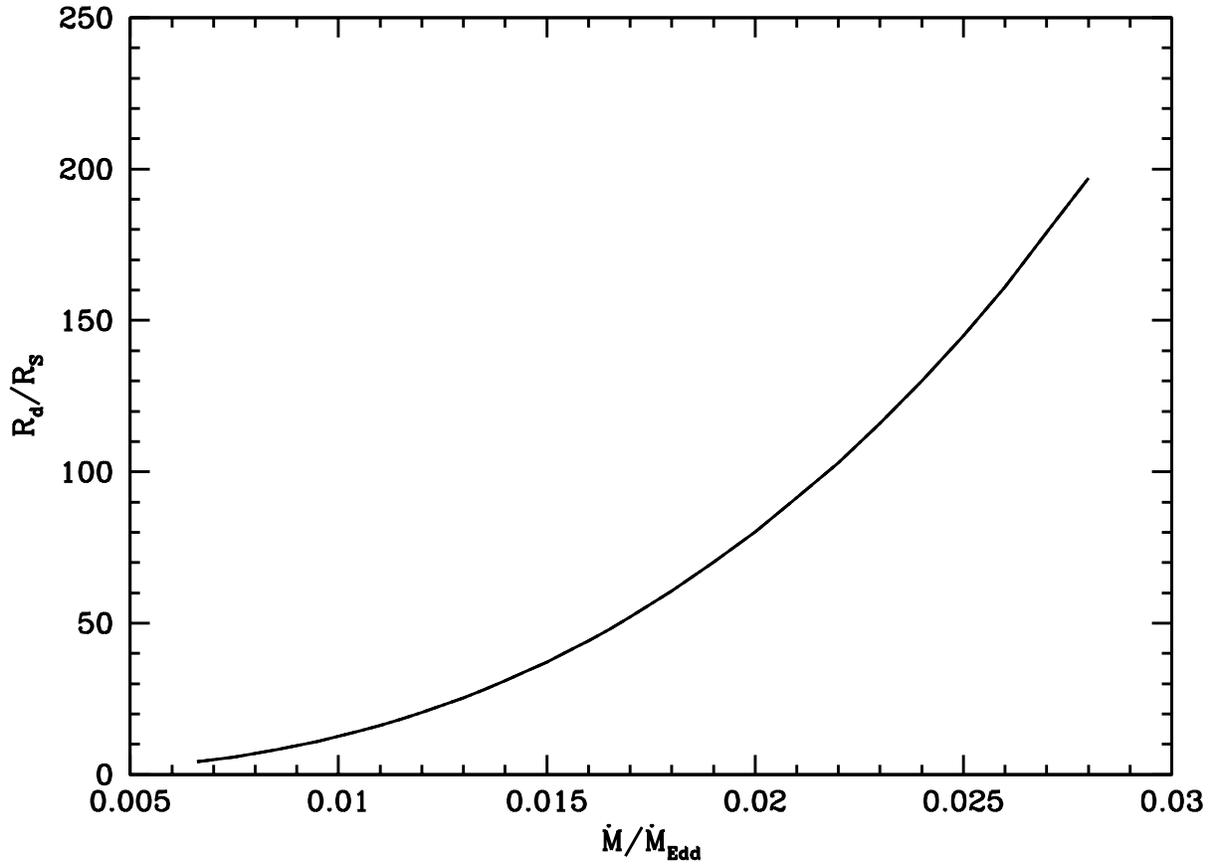}
\caption{\label{f:Rout} The size of the inner disk is shown as a function of accretion rate. The two curves
including irradiation and excluding irradiation overlap, indicating that there is no influence from irradiation.}
\end{figure}

Fig. \ref{f:Te} shows the electron temperature in the corona as a
function of mass accretion rate. There is a marked change in behaviour
at $\dot{m}=0.015$, with irradiation making very little difference
below this point. This is because at low mass accretion rates the
cooling in the corona is dominated by vertical heat conduction rather
than by Comptonisation. Thus increasing the soft photons from the
reprocessed radiation flux makes little difference to the coronal
temperature, so the condensation rate is unaffected.  However, at
higher mass accretion rates, Compton cooling of the corona becomes
important and the inner disk is large. The inner disk subtends a large
angle to the corona radiations, further increasing irradiation.  Thus
the seed photon luminosity largely increases, leading to a decrease in
electron temperature of the corona. This decreases the conductive
heating to the transition layer, and so leads to an increase of the
condensation rate. We show this in Fig. \ref{f:condensation}, defined
as the integrated condensation rate from the outer edge of the inner
disk to the innermost stable circular orbit, ISCO, (so the mass flow
rate through the innermost part of the corona at the ISCO is
$\dot{m}-\dot{m}_{cnd}$).
However, this increased condensation has very little effect on the size of
the residual inner disk (Fig. \ref{f:Rout}) as by this stage the outer edge of the
inner disk is already fairly large.  Irradiation is preferentially of
the inner disk so the increased condensation adds material to the
inner disk rather than increasing its outer radius.

\begin{figure}
\plotone{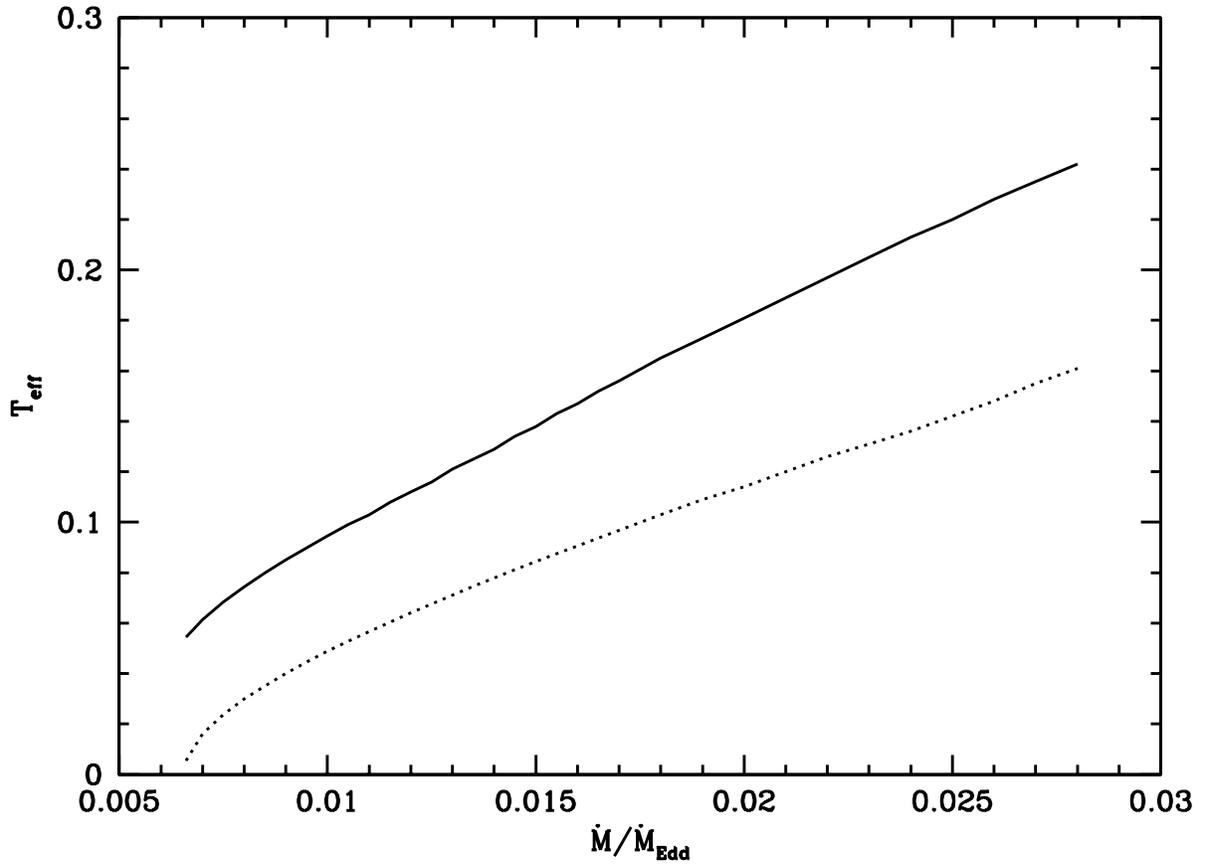}
\caption{\label{f:Teff} The maximal effective temperature of the inner disk heated by accretion and
irradiation is shown as a function of the mass accretion rate. The solid line corresponds to the case with
irradiation and the dotted line for the case without irradiation.}
\end{figure}

While irradiation of the disk makes little difference to the corona at
the lowest mass accretion rates at which the composite geometry can
exist, it does make a difference to the observed emission from the
inner disk because the intrinsic accretion-heated inner disk emission
is extremely dim compared to the corona illumination. Fig.\ref{f:Teff}
shows the inner disk temperature for the irradiated and
non- irradiated case. Irradiation heating means that the disk has a
higher temperature (and luminosity) than expected from its mass
accretion rate alone.  This can be compared to the compilation of
observations of the residual thermal emission in BHXRBs of Reis et al
(2010), where the typical temperature of the soft component is $\sim
0.2$~keV. There is probably a colour temperature correction of $\sim
1.6-1.8$ between the observed and effective temperature (Shimura \&
Takahara 1995), so the data probably correspond to an effective
temperature of $\sim 0.10-0.15$~keV, matching well with the
predictions of the irradiated inner disk for low mass accretion rates
($\dot{m}\la 0.02$).

The enhanced luminosity of the disk also impacts on the observed
spectrum of the corona. At the lowest luminosity the coronal radiation
is dominated by bremstrahlung rather than Compton scattering, so the
(small) increase in seed photon flux makes little difference to the
spectrum. However, Comptonisation increases in importance as the mass
accretion rate increases, and is comparable to bremstrahlung from the
transition layer at $\dot m=0.015$. Above this, Comptonisation is
dominant, and the ratio of (reprocessed) disk to coronal emission sets
the slope of the Comptonised spectrum so that the larger this ratio,
the softer the spectrum (Haardt \& Maraschi 1993).

Most of the spectra in Reis et al (2010) show a weak soft
component co-existing with hard spectra, with $\Gamma<1.8$. From
Compton cooling models, both numerical ({\sc eqpair}: Coppi 1991) and
analytic (Pietrini \& Krolik 1995) require that the ratio
$L_d/L_{c}<0.3$. This ratio can be estimated analytically as a
function of the covering fraction of the disk, $f$, and its albedo,
$a$, so that $L_d/L_{c}=f(1-a)/\[1-f(1-a)\]$ (Haardt \& Maraschi
1993).  This is a lower limit as there should be some intrinsic
dissipation from the material accreting through the disk. Hence
$\Gamma<1.8$ (equivalent to $0.3>L_d/L_{c}>f(1-a)/\[1-f(1-a)\])$
requires $f<0.27$ for $a=0.15$. Such a low covering fraction requires
that the residual inner disk is rather small, corresponding to
$R_{out}=13.5R_s$. Such a small disk is only possible at very low mass
accretion rates, so there is very little range in $\dot{m}$ over which
the residual disk can co-exist with a hard spectrum.

However, this conclusion is very dependent on the assumed albedo. A
more reflective disk (so that less of
the irradiating luminosity is reprocessed) can give harder spectra for
larger inner disks. However, hard spectra peak at high energies, where
Compton downscattering means that the energy cannot be completely
reflected. The leads to a maximum possible albedo of $a\sim 0.6$,
corresponding to a covering fraction $f<0.58$. This estimate appears not to constrain the disk size. We note that the disk accretion could contribute a luminosity comparable with the reprocessed disk luminosity in the case of a large albedo, $a=0.6$. Including this effect, the covering factor required to produce spectra $\Gamma<1.8$ is smaller.

\begin{figure}
\plotone{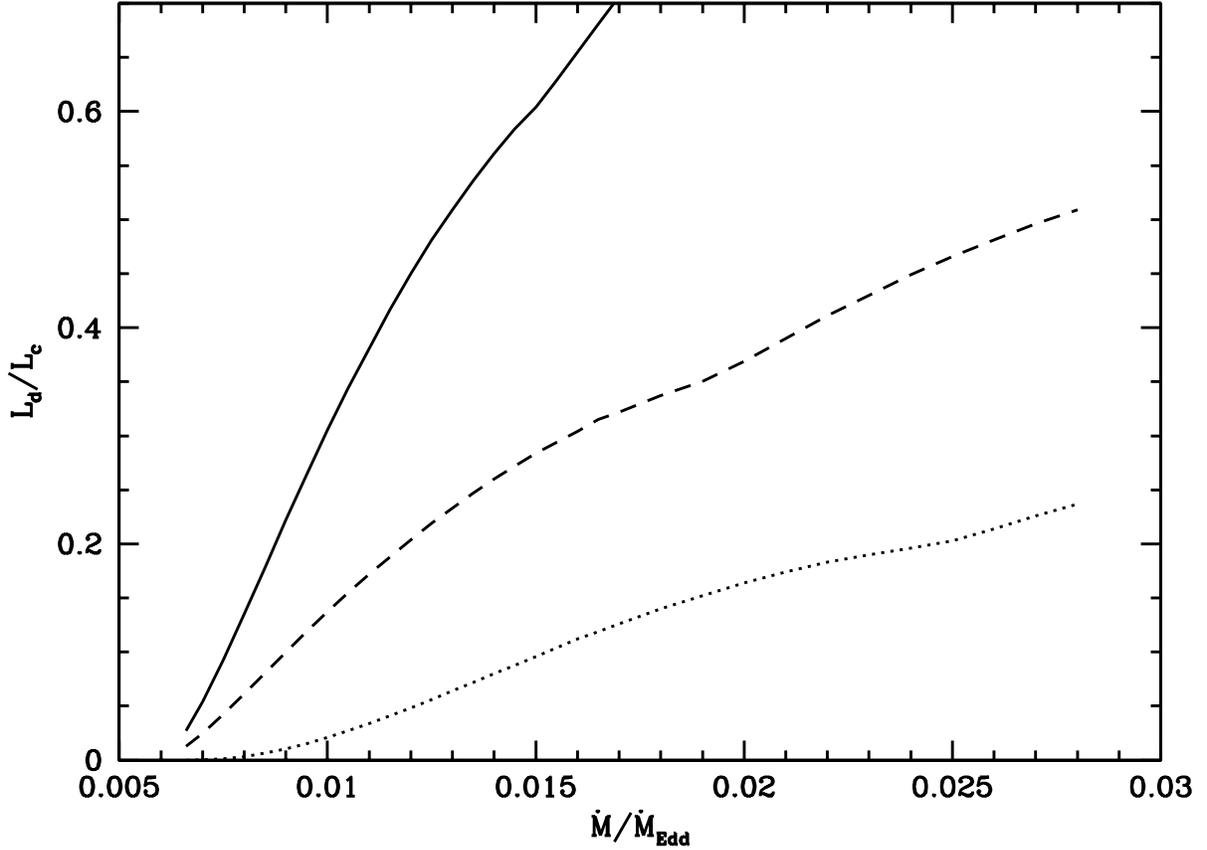}
\caption{\label{f:ratioL} The ratio of the luminosity from the disk relative to the corona is shown as a
function of the mass accretion rate.  The solid line corresponds to the case with irradiation and the dotted
line to the case without irradiation. Here $L_{c}$ denotes the emergent coronal luminosity where the
fraction absorbed by the inner disk is deducted from the coronal emission. The
luminosity ratio is significantly increased by the irradiation, however, it steeply drops to zero when the
accretion rate decreases to the critical value of $\sim 0.006$. The influence of the albedo is shown by
the dashed line, where $a = 0.6$ is assumed.  The effect of irradiation is much reduced if a larger
fraction of the irradiation flux is reflected at the disk surface.}
\end{figure}

We confirm these estimates with the numerical results of $L_d/L_c$
from our models. This ratio is shown in Fig. \ref{f:ratioL} for the
representative albedos of 0.15 and 0.6, and the requirement that the
ratio is less than 0.3 in order to produce hard spectra ($\Gamma<1.8$)
directly constrains $\dot{m}< 0.01$ or $0.016$ for $a=0.15$ and 0.6,
respectively.

Thus the model can indeed produce a residual inner disk temperature and
hard coronal luminosity to match the observations of the small thermal
component seen together with a hard X-ray continuum, but only for a
limited range in mass accretion rate of $0.006<\dot{m}<0.01$ for
mostly neutral reflection or $0.006<\dot{m}<0.016$ for a highly
ionised disk. Thus the model predicts that the composite geometry
(outer disk, coronal gap, with mass accretion rate in the innermost
regions split between a weak disk and coronal flow) can only be seen
over a range in luminosity of at most factor $\sim 7$ assuming an
efficiency of the coronal flow, $\eta \propto \dot m$.  Even this
maximum range in luminosity is slightly smaller compared to the
observations of Reis et al (2010). These show the thermal component
together with hard spectral index ($\Gamma<1.8$, so J1650-500 with
$\Gamma=2.1$ excluded) over a range of about a factor 10 in
$L_X/L_{\rm Edd}$ (more if XTE J1118+480 is included), but we note
there are substantial uncertainties in system parameters, so this can
be consistent with the data.  A bigger discrepancy is that
the mass accretion rate of $\dot m=0.006-0.016$ in the model corresponds to a luminosity of
$L/L_{Edd} \approx 1-9\times 10^{-3}$ (assuming a self-similar ADAF with energy conversion efficiency $\eta=0.1\times{\dot m\over 0.03})$,
somewhat smaller than that seen in the data
which cluster around $L/L_{Edd} \la
5-9\times 10^{-3}$ (Reis et al 2010).
However, the scaling from mass accretion rate to luminosity
is dependent on the model
parameters chosen, particularly on $\alpha$ (Esin et al 1998).

At higher mass accretion rates, the spectrum softens strongly as the
increased seed photon flux leads to increased Compton cooling of the
corona. This means that the {\em spectral} transition is triggered
earlier than expected from the statement that the disk should exist at
all radii. Models with $\dot{m}\sim 0.02$ do not have a continuous
disk. There is still a coronal gap, but the spectrum is formed mainly
in the innermost regions, and at these radii the disk covers almost
half the sky as seen from the corona. Comptonised emission calculated
from such a composite accretion flow with electron temperature and
optical depth determined by our model for $\dot m\sim 0.02$ yields
a photon index $\Gamma\sim 2$. Thus the transition from the low/hard
to intermediate state occurs at $\dot m\sim 0.02$ or  $L/L_{Edd} \sim
1.3\times 10^{-2}$.

\section{Conclusions}

We investigate the influence of coronal radiation on the properties of
an existing residual inner disk.  For accretion rates close to the hard-to-soft transition rate ($\dot m\sim 0.03$),
the disk is only truncated in a very
narrow radial extent by mass evaporation.  In this case, the disk is
described by a geometrically thin inner region, a narrow geometrically
thick intermediate region, and a geometrically thin outer region. Here
we find that including irradiation results in efficient
Compton cooling and condensation of coronal gas to the inner disk,
with the emergent spectrum much softer than
predicted without reprocessing.  This means that the transition from
the intermediate to low/hard state occurs at a lower mass
accretion rate, at
$\dot m\sim 0.016-0.02$ for the specific
parameters assumed here, rather than at $\dot m\sim 0.03$ without
irradiation.

Below this, evaporation from the outer region results in an increased
radial extent of the ADAF region and a smaller inner disk. The
resulting geometry leads to a greater fraction of photons which can
escape from the ADAF without irradiating the inner disk. Therefore, at
sufficiently low accretion rates ($\dot m \la 0.015$),
the irradiation of the inner disk is
small and the resulting Compton cooling does not significantly
influence the dynamical process in the corona. Consequently, the
condensation rate is not affected, while the inner disk is heated to a
higher effective temperature by irradiation.

Thus a weak, inner disk can exist in the low/hard state even if
irradiation is taken into account. The ratio of luminosities emergent
from the disk and corona decrease with decreasing mass accretion
rates, until $\dot m\sim 0.006$ at which mass condensation no longer
occurs and the inner disk cannot be sustained.  Our results are in
broad agreement with recent X-ray observational results that systems
with a weak disk can exist in the low/hard spectral state (Reis et al.
2010).

\acknowledgments
B.F.L. thanks the hospitality of Department of Physics, Durham
University, where this work was initiated.  Financial support for this
work is provided by the National Natural Science Foundation of China
(11033007 and 10773028) and by the National Basic Research
Program of China-973 Program 2009CB824800. In addition,
R.E.T. acknowledges support from the Theoretical Institute for
Advanced Research in Astrophysics (TIARA) in the Academia Sinica
Institute of Astronomy \& Astrophysics.

{}

\end{document}